\documentstyle[prd,floats,aps,epsfig]{revtex}

\begin{document}

\input{epsf.sty}

\draft

\twocolumn[\hsize\textwidth\columnwidth\hsize\csname
@twocolumnfalse\endcsname

\title{Black Hole Excision for Dynamic Black Holes}

\author{
Miguel Alcubierre${}^{(1)}$,
Bernd Br\"ugmann${}^{(1)}$,
Denis Pollney${}^{(1)}$,
Edward Seidel${}^{(1,2)}$,
Ryoji Takahashi${}^{(1)}$}

\address{
$^{(1)}$ Max-Planck-Institut f{\"u}r Gravitationsphysik,
Am M\"uhlenberg 1, D-14476 Golm, Germany
}
\address{
$^{(2)}$ National Center for Supercomputing Applications,
Beckman Institute, 405 N. Mathews Ave., Urbana, IL 61801
}

\date{\today; AEI-2001-021}

\maketitle


\begin{abstract}
  We extend previous work on 3D black hole excision to the case of
  distorted black holes, with a variety of dynamic gauge conditions
  that respond naturally to the spacetime dynamics.  We show that in
  evolutions of highly distorted, rotating black holes, the
  combination of excision and the gauge conditions we use is able to
  drive the coordinates to a frame in which the system looks almost
  static at late times.  Further, we show for the first time that one
  can extract accurate waveforms from these simulations, with the full
  machinery of excision and dynamic gauge conditions.  The evolutions
  can be carried out for long times, far exceeding the longevity and
  accuracy of better resolved 2D codes.
\end{abstract}


\narrowtext

\vskip2pc]


{\em Introduction.} The long term simulation of black hole (BH)
systems is one of the most challenging and important problems in
numerical relativity.  For BHs, the difficulties of accuracy and
stability in solving Einstein's equations numerically are exacerbated
by the special problems posed by spacetimes containing singularities.
At a singularity, geometric quantities become infinite and cannot be
handled easily by a computer.

Traditionally, in the 3+1 approach the freedom in choosing the slicing
has been used to slow down the approach of the time slices towards the
singularity (``singularity avoidance''), while allowing them to
proceed outside the BH \cite{Smarr78b}.  Singularity avoiding slicings
are able to provide accurate evolutions, allowing one to study BH
collisions and extract waveforms~\cite{Alcubierre00b}, but only for
limited evolution times.  Combining short full numerical evolutions
with perturbation methods, one can even study the plunge from the last
stable orbit of two BHs~\cite{Baker01b}.  But a breakthrough is
required to push numerical simulations far enough to study orbiting
BHs, requiring accurate evolutions exceeding time scales of $t \approx
100M$.  In 3D, traditional approaches have not been able to reach such
time scales, even in the case of Schwarzschild BHs.  Characteristic
evolution codes, on the other hand, are well-adapted to the long-term
evolution of single black holes \cite{Gomez98a}, but here we
concentrate on non-characteristic methods for their ease of
generalization to multiple black holes.

A more promising approach involves cutting away the singularity from
the calculation (``singularity excision''), assuming it is hidden
inside an apparent horizon (AH) \cite{Thornburg87,Seidel92a}.  This
work has been progressing, from early spherical proof of
principle~\cite{Seidel92a} to recent 3D
developments~\cite{Gomez98a,Cook97a,Alcubierre00a,Brandt00,Kidder01a}.
However, beyond a few spherical test cases~\cite{Anninos94c,Daues96a},
excision has yet to be used in conjunction with live gauge conditions
designed to respond to both the dynamics of the BH and the coordinate
motion through the spacetime.

In this paper we extend recent excision work~\cite{Alcubierre00a} to
the case of distorted, dynamic BHs in 3D, using a new class of gauge
conditions.  These gauge conditions not only respond naturally to the
true spacetime dynamics, but also {\em drive the coordinates to
a frame where the system looks almost static at late times}.  We show
that not only are the evolutions accurate as indicated by the mass
associated with the apparent horizon, but also that very accurate
waveforms can be extracted, even when the waves carry only a tiny
fraction of the energy of the spacetime.  We also show that the 3D
evolutions of dynamic BHs we are now able to perform, are superior, in
terms of accuracy, stability, and longevity, to previous 3+1 BH
simulations, whether carried out in full 3D or even when restricted to
2D. These results indicate that BH excision can be made to work under
rather general circumstances, and can significantly improve the length
of the evolutions, and the accuracy of the waveforms extracted, which
will be crucial for gravitational wave astronomy.


{\em Initial Data.} For this paper we consider single distorted BH
spacetimes~\cite{Brandt94a,Brandt94c} that have been used to model the
late stages of BH coalescence~\cite{Camarda97c,Baker99a}.
Following~\cite{Brandt94a,Brandt94c}, the initial 3-metric
$\gamma_{ab}$ is chosen to be
\begin{equation}
ds^2=\Psi^4 \left[ e^{2 q} \left(d \eta^2+d\theta^2 \right)
+\sin^2\theta\, d\phi^2 \right],
\label{eq:mymet}
\end{equation}
where the ``Brill wave'' function $q$ is a general function of the
spatial coordinates, subject to certain regularity and fall off
restrictions, that can be tailored to produce very distorted 3D BHs
interacting with nonlinear waves.  The radial coordinate $\eta$ is
logarithmic in the cartesian radius $r$.  There are two classes of
data sets used here corresponding to even- and odd-parity distortions.
The even-parity data have vanishing extrinsic curvature, while the
cases containing an odd-parity component have nontrivial extrinsic
curvature $K_{ij}$.  As shown in~\cite{Brandt97c,Brandt97a}, these
distorted BH data sets can include rotation as well, corresponding to
spinning, distorted BHs that mimic the early merger of two orbiting
BHs.  Hence they make an ideal test case for the development of our
techniques.  We leave the details of the construction of these BH
initial data sets to Refs.~\cite{Brandt97c,Brandt97a}.  An important
point that we wish to emphasize is that such data are {\em not} of the
Kerr-Schild form with ingoing coordinates at the horizon.  That
form of initial data sets has been recently advocated
since it is not conformally flat~\cite{Matzner98a} and is well
adapted to inward propagation of quantities at the horizon.


{\em Evolution and Excision Procedures.} Our simulations have been
performed using what we refer to as the ``BSSN'' version of the 3+1
evolution
equations~\cite{Shibata95,Baumgarte99,Alcubierre99b,Alcubierre99e},
which we have found to have superior stability properties when
compared to standard formulations.  As detailed
in~\cite{Alcubierre99b,Alcubierre99e,Alcubierre99d}, we actively force
the trace of the conformal-traceless extrinsic curvature
$\tilde{A}_{ij}$ to remain zero, and we use the independently evolved
``conformal connection functions'' $\tilde\Gamma^i$ only in terms
where derivatives of these functions appear. All the simulations
described below have been performed using a 3-step iterative
Crank-Nicholson scheme and a radiative (Sommerfeld) outer boundary
condition.  We refer the reader to Ref.~\cite{Alcubierre99d} for the
details of the numerical implementation.

We use the simple excision approach described in~\cite{Alcubierre00a}.
Our algorithm is based on the following ideas: (a) Excise a {\em cube}
contained inside the AH that is well adapted to cartesian coordinates;
(b) Use a simple boundary condition at the sides of the excised cube:
copying of time derivatives from their values one grid point out along
the normal directions; (c) Use centered (non-causal) differences in
all terms except for advection terms on the shift (terms of the form
$\beta^i \partial_i \,\,$).  For these terms we use second order
upwind along the shift direction.  These simplifications in excision
reduce the complexity in the algorithm, avoid delicate interpolation
issues near the excision boundary, and have allowed us to make rapid
progress. Currently, the method is implemented for non-moving excision
regions, although they are allowed to grow. One can hope that
colliding black holes can be treated even with this restriction
through the use of co-moving coordinates. A more detailed description
of our excision algorithm can be found in Ref.~\cite{Alcubierre00a}.



{\em Gauge Conditions.} For the lapse we use a hyperbolic
slicing condition motivated by the Bona-Mass\'{o}
family of slicing conditions~\cite{Bona94b},
\begin{equation}
\partial_t \alpha = - \alpha^2 \, f \left(\alpha \right) \,
(K - K_0),
\label{eq:modifiedlapse}
\end{equation}
where $K$ is the trace of the extrinsic curvature, $K_0$ is its value
in the initial data, and $f$ is a (positive) function of $\alpha$
which we specify below.  With this condition, the lapse will evolve as
long as $\alpha^2 f(\alpha)$ and $K-K_0$ are non-vanishing.


For the shift $\beta^i$ we have considered families of elliptic,
parabolic, and hyperbolic conditions that relate the shift with the
evolution of the conformal connection functions $\tilde\Gamma^i$. We
obtain parabolic and hyperbolic shift conditions by making either
$\partial_t \beta^i$ or $\partial^2 _t \beta^i$ proportional to the
elliptic operator for $\beta^{i}$ contained in the ``Gamma freezing''
condition $\partial_t \tilde\Gamma^k=0$ (see
Ref.~\cite{Alcubierre99d}), itself closely related to the well known
minimal distortion family~\cite{Smarr78b}.  Elliptic conditions have
the disadvantage of requiring boundary data at the excision region
where it is difficult to know what to impose, while parabolic
conditions force a strong restriction on the stability of the
differencing scheme: $\Delta t \propto (\Delta x)^2$ (this is true for
explicit schemes, implicit schemes have no such restriction).  We have
then concentrated on hyperbolic conditions of the form
\begin{equation}
\partial^2_t \beta^i = \zeta \, \partial_t \tilde\Gamma^i
- \xi \, \partial_t \beta^i \, ,
\label{eq:hyperbolicGammadriver}
\end{equation}
where $\zeta$ and $\xi$ are positive functions.
We call such evolution conditions for the 
shift hyperbolic ``driver'' conditions
(see~\cite{Balakrishna96a}).

In the spirit of the puncture method for
evolutions~\cite{Bruegmann97}, we use a BSSN scheme with the usual
time-dependent conformal factor $e^{4\phi}$ and an additional
time-independent conformal factor $\Psi^4$ that comes from the initial
data. In our examples we use $\zeta = k / \Psi^4$, where $k$ is a
positive constant.  The division by $\Psi^4$ helps to slow down the
evolution of the shift in the vicinity of the black hole.  We have
found it important to add a dissipation term with a constant
coefficient $\xi$ in order to reduce some initial oscillations in the
shift.  Notice that in contrast with $k$, the coefficient $\xi$ is not
dimensionless (it has dimensions of inverse length), so in practice we
rescale it using the total mass of the system.  Experience has shown
that by tuning the value of $\xi$ we can almost freeze the evolution
of the system at late times.

The parameters used for all simulations described below are: $\alpha$
is given by Eq.~(\ref{eq:modifiedlapse}), with $f=2/\alpha$;
$\beta^{i}$ is given by Eq.~(\ref{eq:hyperbolicGammadriver}) with
$\zeta=0.75/ \Psi^4, \xi=3/M$ ($M$ is the initial ADM mass of the
system).  As initial conditions we take $\alpha=1$, $\beta^{i}=0$,
$\partial_t \alpha = \partial_t \beta^{i} = 0$, except in one case
mentioned below where we perform a single maximal slicing solve to
obtain a more appropriate initial lapse.  Given these initial
conditions, we let the gauge conditions take care of the rest.  We use
the same gauge parameters for all results in this paper, whether they
are applied to Schwarzschild, distorted, or rotating BHs, showing the
strength and generic nature of these conditions.


\begin{figure}[t]
\epsfxsize=85mm \epsfysize=80mm \epsfbox{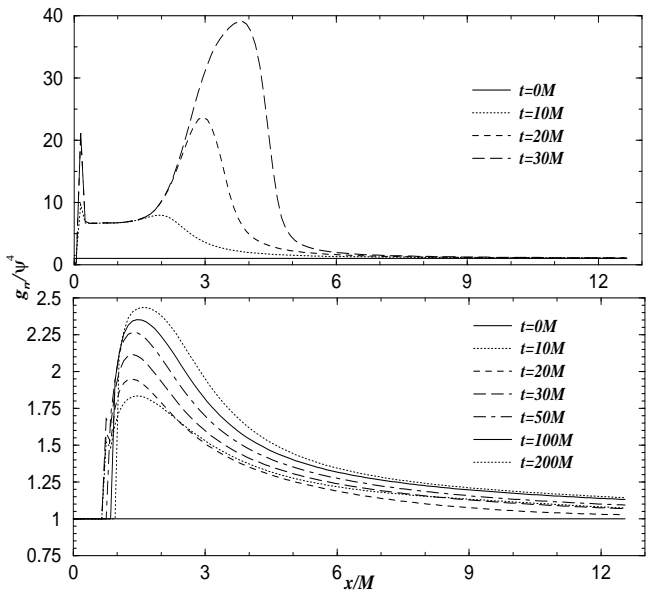} 
\caption{We show the evolution of the radial metric function
  $g_{rr}/\Psi^{4}$ for a Schwarzschild BH along the $x-$axis,
  constructed from the cartesian components.  The upper panel shows
  the grid stretching in the metric for singularity avoiding slicing
  with vanishing shift and no excision, while the lower panel shows
  the metric for the new gauge conditions with an excision box inside
  a sphere of radius $1M$.  Note the difference in the vertical
  scales.  Without shift and excision the metric grows out of control,
  while with shift and excision a peak begins to form initially but
  later freezes in as the shift drives the metric to a static
  configuration (note the time labels).}
\label{fig:schwarz1}
\end{figure}

\begin{figure}
\epsfxsize=85mm \epsfysize=80mm \epsfbox{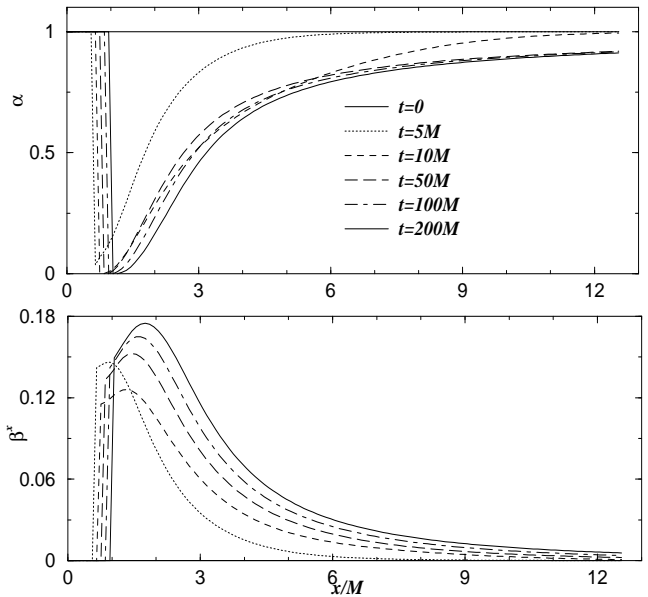}
\caption{We show the lapse and shift for the excision evolution of a
  Schwarzschild BH. After around 10M, the lapse and shift freeze in as
  the metric is driven to a static configuration. The size of the
  excision box was allowed to grow with the change in the coordinate
  location of the AH.}
\label{fig:schwarzgauge}
\end{figure}

{\em Results.} The first example we show is Schwarzschild, written in
the standard isotropic coordinates used in many BH evolutions.  Note
that with this initial data and our starting gauge conditions, the BH
should evolve rapidly.  If $\alpha$ and $\beta^{i}$ were held fixed at
their initial values, the slices would hit the singularity at $t=\pi
M$.  Instead, $\alpha$ and $\beta^{i}$ work together with excision
to rapidly drive the coordinates to a frame where the system looks
essentially static, corresponding to the true physical situation.

In Fig.~\ref{fig:schwarz1} we show the radial metric function
$g_{rr}/\Psi^{4}$ vs.  time.  The grid covers an octant with $128^3$
points ($\Delta x = 0.2$, $M=2$).  Appropriate symmetry conditions are
applied on the faces of the octant for the different dynamical
variables. We have checked that removing the octant symmetry (while
using a lower resolution) does not change the results for the
evolution times reported here (in particular no instabilities were
encountered, cmp.~\cite{Alcubierre00a}).  Notice that the metric
begins to grow, as it does without a shift, but as the shift builds up
the growth slows down significantly. At this stage, the system is
effectively static, even though we started in the highly dynamic
isotropic coordinates.  We also show the time development of $\alpha$
and $\beta^{r}$ in Fig.~\ref{fig:schwarzgauge}, which evolve rapidly
at first but then effectively freeze, bringing the whole system to an
almost static state by $t=10M$. The evolution of the metric and gauge
variables then proceeds only very slowly with time until the
simulation is stopped well after $t=200M$.  The decision to stop the
code at this time is simply a CPU time consideration, but we notice
that in this and all following examples the code is stopped once all
the interesting dynamics have finished.

\begin{figure}[t]
\epsfxsize=85mm \epsfysize=50mm \epsfbox{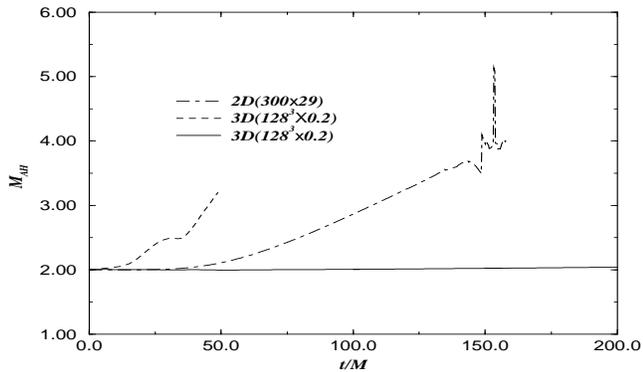}
\caption{
  The solid line shows the development of the AH mass $M_{AH}$,
  determined through a 3D AH finder, for the simulation of a
  Schwarzschild BH shown above, while the dashed lines show the AH
  mass obtained using 2D and 3D codes with no shift and no excision.
  The 2D code crashes at $t \simeq 150M$, the 3D run without shift
  crashes at $t \simeq 50M$, while the 3D run with shift and excision
  reaches an effectively static state and the error remains less than
  a few percent even after $t=200M$.}
\label{fig:schwarz2}
\end{figure}

Figure~\ref{fig:schwarz2} shows the AH mass \mbox{$M_{AH} = \sqrt{{\rm
      Area}_{AH}/16\pi}$}, determined with a 3D AH
finder~\cite{Alcubierre98b}.  For comparison we also show the value of
$M_{AH}$ for the 3D run without shift, and for a highly resolved 2D
simulation with no shift and no excision \cite{Brandt94c}. The 2D code
uses maximal slicing, so the coordinate time $t$ refers to different
slices, but the slicings turn out to be even more similar than is to
be expected from Eq.~(\ref{eq:modifiedlapse}).  While the 3D
simulation with shift and excision continues well beyond $t=200M$, the
2D result becomes inaccurate and the code crashes due to axis
instabilities by $t=150M$, and the 3D run without shift crashes
already by $t=50M$. Notice that in the 2D case, after around $t=35M$,
$M_{AH}$ grows rapidly due to numerical errors associated with grid
stretching.  With excision and our new gauge conditions, the 3D run
has less than a few percent error by $t=200M$, while the 2D case has
more than 100\% error before it crashes at $t \approx 150M$.  For the
excision run, notice also that while there is some initial evolution
in the metric and the coordinate size of the AH (see
Figs.~\ref{fig:schwarz1} and \ref{fig:schwarzgauge}) the AH mass
changes very little.

Next, we turn to a truly dynamic, even-parity distorted BH. This
system contains a strong gravitational wave that distorts the BH,
causing it to evolve, first nonlinearly, and then oscillating at its
quasi-normal frequency, finally settling down to a static
Schwarzschild BH.  This provides a test case for our techniques with
dynamic, evolving BH spacetimes, and allows us to test our ability to
extract gravitational waves with excision for the first time.  In this
case, in the language of~\cite{Brandt97c}, we choose the Brill wave
parameters to be $Q_0=0.5$, $\eta_0=0$, $\sigma=1$, corresponding to a
highly distorted BH with \mbox{$M=1.83$}.  Just as before, we use a
grid that covers one octant, with $128^3$ points and $\Delta x = 0.2$

\begin{figure}[t]
\epsfxsize=85mm \epsfysize=50mm \epsfbox{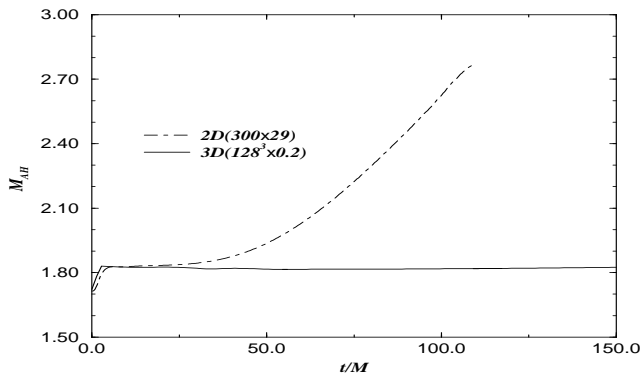}
\caption{We show the AH masses $M_{AH}$ for a BH with even-parity
  distortion for the 2D (no excision, no shift) and 3D (excision,
  shift) cases.  The 3D result continues well past $150M$, while the
  2D result becomes very inaccurate and crashes by $t=100M$.}
\label{fig:even1}
\end{figure}

\begin{figure}
\epsfxsize=85mm \epsfysize=50mm \epsfbox{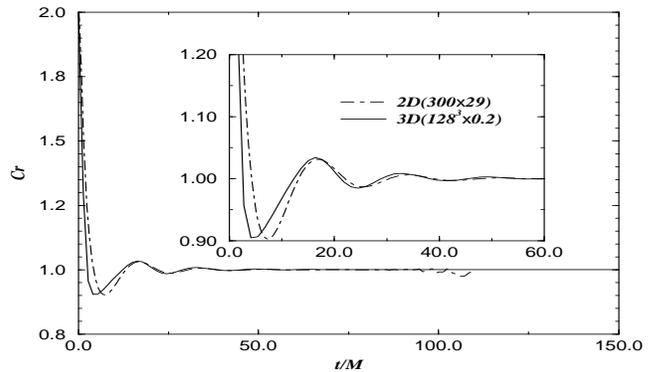}
\caption{We show the ratio of the polar and equatorial circumferences
  as a measure of the dynamics of the AH for the BH with even-parity
  distortion .}
\label{fig:evencr}
\end{figure}

In Fig.~\ref{fig:even1} we show the AH mass $M_{AH}$ as a function of
time for the distorted BH simulations carried out in both 2D and 3D.
$M_{AH}$ grows initially as a nonlinear burst of gravitational waves
is absorbed by the BH, but then levels off as the BH goes into a
ring-down phase towards Schwarzschild.  Fig.~\ref{fig:evencr} shows
the proper polar circumference of the AH divided by its
proper equatorial circumference. This ratio allows an estimate of the
size of the local dynamics during the run.  Notice how the horizon
starts far from spherical (with a ratio close to 2), it later
oscillates from prolate to oblate and back again, and finally settles
on a sphere (with a ratio of 1).

In the 3D case, the gauge conditions and excision quickly drive the
metric to an almost static configuration, as the system itself
settles towards a static Schwarzschild BH. The evolution is terminated
at around $t= 160M$.  To our knowledge, distorted BHs of this type
have never been evolved for so long, nor with such accuracy, in either
2D or 3D.  By comparison, in the more highly resolved 2D case with
zero shift and no excision, the familiar grid stretching effects
allowed by the gauge choice lead to highly inaccurate evolutions after
some time with the error in $M_{AH}$ again approaching 100\% when the
code finally crashes at $t \approx 100M$.

\begin{figure}[t]
\epsfxsize=85mm
\epsfysize=80mm
\epsfbox{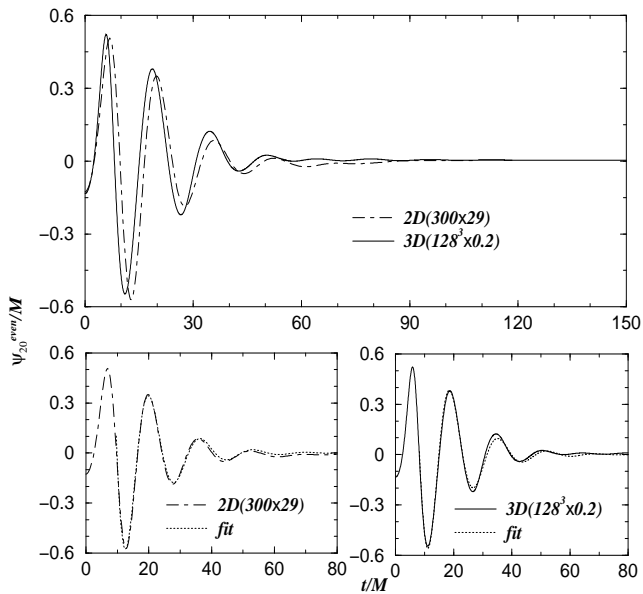} 
\caption{The solid line shows the $\ell=2, m=0$ waveform extracted at
  a radius of $5.45M$ for the even-parity distorted BH described in
  the text, while the dashed line shows the result of the same
  simulation carried out in the 2D code. We also show a fit to the two
  lowest QNM's of the BH for 2D and 3D separately, using numerical
  data from $t=9M$ to $t=80M$.  }
\label{fig:even2}
\end{figure}

In Fig.~\ref{fig:even2}, we show the results of extracting waves from
the evolution of this highly distorted, excised BH. Using the standard
gauge-invariant waveform extraction technique, the Zerilli function is
shown for both the 2D and 3D simulations discussed above.  There is a
slight but physically irrelevant phase difference in the two results
due to differences in the slicing; otherwise the results are
remarkably similar (the waves are extracted at the same Schwarzschild
radius in both cases).  This shows conclusively that the excision and
live gauge conditions do not adversely affect the waveforms, even if
they carry a small amount of energy (around $10^{-3}M_{ADM}$ in this
case).

\begin{figure}[t]
\epsfxsize=85mm
\epsfysize=50mm
\epsfbox{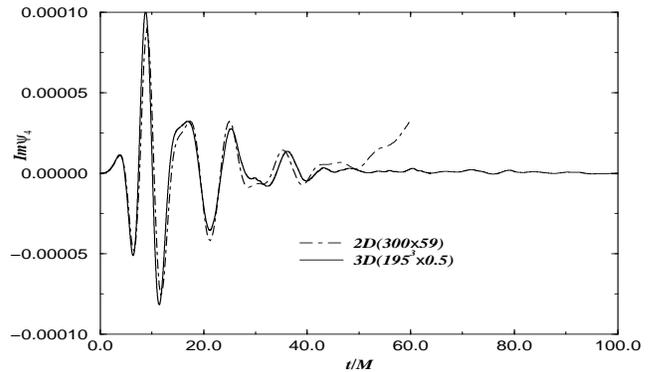}
\caption{The solid line shows the imaginary part of $\psi_4$ computed
at $r=3.94M$ and $\theta=\phi=\pi/4$ for a rotating distorted BH
obtained from our $3D$ code with excision, while the dash line shows
the same quantity computed using a $2D$ code (this simulation crashes
at $t \simeq 60M$).}
\label{fig:rotor3d}
\end{figure}

We now turn to a rather different type of distorted BH, including
rotation and general even and odd-parity distortions.  In the language
of Ref.~\cite{Brandt97c}, the parameters for this simulation are
$Q_0=0.5$, $\eta_0=0$, $\sigma=1$, \mbox{$J=35$}, corresponding to a
rotating distorted BH with \mbox{$M=7.54$} and an effective rotation
parameter $a/M = 0.62$. Previously, such data sets could be evolved
only to about $40M$ \cite{Baker99a}.  For the purposes of this paper
we have chosen an axisymmetric case so that we can compare the results
to those obtained with a 2D code. Since this example is much more
demanding, we have found it important in order to increase the
accuracy of our runs to perform a single initial maximal solve to
reduce the initial gauge dynamics.  The symmetries of this example are
now not consistent with the evolution of just one octant. However, we
still have reflection symmetry on the $z=0$ plane, so we evolve only
the positive $z$ half of the domain.  The grid used in this case has
$199^2\times100$ points and $\Delta x = 0.4$.  The gauge conditions
work well even in the presence of rotation: the shift drives the
evolution to an almost static state as the system itself
settles down to a {\em Kerr} BH.  The metric functions (not shown)
evolve in a similar way to those shown before, essentially freezing at
late times.  In Fig.~\ref{fig:rotor3d}, we show the extracted
waveforms, now computed using the imaginary part of the Newman-Penrose
quantity $\Psi_{4}$ (e.g.~\cite{Baker01b}), which includes
contributions from all $\ell-$modes at the same time.  The results
from the 2D and 3D codes agree very closely, except for a slight phase
shift due to slicing differences, until the 2D code becomes inaccurate
and crashes.  The 3D simulation continues well beyond this point, and
is terminated at $t=120M$.

{\em Conclusions.} We have extended recently developed 3D BH excision
techniques, using a new class of live gauge conditions that {\em
dynamically drive} the coordinates to a frame where the metric looks
essentially static at late times, when the system itself settles to a
stationary Kerr BH.  Our techniques have been tested on highly
distorted, rotating BHs, and are shown to be very robust. For the
first time, excision is tested with wave extraction, and waveforms are
presented and verified.  The results are shown to be more accurate,
and much longer lived, than previous 3D simulations and even better
resolved 2D simulations of the same data.  Such improvements in BH
excision are badly needed for more astrophysically realistic BH
collision simulations, which are in progress and will be reported
elsewhere.

\noindent {\bf Acknowledgements.} This work was supported by AEI.
Calculations were performed using the Cactus code at AEI, NCSA, PSC,
and RZG.


\bibliographystyle{prsty}
\bibliography{bibtex/references}

\end{document}